\documentclass[12pt,thmsa]{article}
\usepackage{amsmath}

\input tcilatex
\begin{document}

\title{Hidden Galilean symmetry, conservation laws and emergence of spin
current in the soliton sector of chiral helimagnet}
\author{I.G. Bostrem$^1$, J.Kishine$^2$, R.V. Lavrov$^1$, A.S. Ovchinnikov$%
^1 $ \\
$^1$Department of Physics, Ural State University, \\
Ekaterinburg, 620083, Russia \\
$^2$Faculty of Engineering, Kyushu Institute of Technology,\\
Kitakyushu 804-8550,Japan}
\date{\today }
\maketitle

\section{Introduction}

The multidisciplinary field of spintronics has attracted keen scientific
interest from the viewpoints of both fundamental and applied physics.\cite%
{Zutic04} The key notion there is how to carry the intrinsic spin
magnetization in a well defined fashion.\cite{Sloncew,Berger,Rashba05} The
current induced motion of a magnetic domain involves a torque transfer
process from the conduction electrons which generates a sliding mode of the
domain wall, i.e. a motion without distortion and without tilting and
twisting\cite{Bazaliy,Barnes}. Naturally speaking, the spin current is
understood as the deviation of the spin projection from its equilibrium
value.

Quite recently, we proposed a new way to generate a spin current in the
so-called chiral helimagnets where the axial coupling between the
neighboring spins called Dzyaloshinskii-Morya (DM) coupling fixes the left-
or right-handedness of the helical spin arrangement over macroscopic scales.%
\cite{BKO1} By applying the static magnetic field perpendicular to the
helical axis, the magnetic kink crystal (chiral soliton lattice) is formed%
\cite{Dzyaloshinskii,Dzyal1} and the inertial motion of the kinks (Galilean
boost)\ triggers of the transport of magnon density (spin current) inside
the magnetic background\cite{BKO1}. Starting with the effective lattice
Hamiltonian and its continuum version, we wrote down the Lagrangian
including the appropriate Berry phase term and introduced the sliding
solution of the soliton lattice by inspection. Then, we constructed the
canonical Hamiltonian for the collective coordinate variables which describe
the center-of-mass motion of the whole soliton lattice, based on the Dirac's
prescription for the singular Lagrangian. After these procedures, we
succeeded in describing the internal motion of the kink crystal state. The
most important notion is that the \textquotedblleft spin
phase\textquotedblright directly comes up in the observable effects through
the soliton lattice formation. In our mechanism, the current is carried by
the moving magnetic kink crystal, where the linear momentum has a form, $%
P=2\pi S\mathcal{Q}+M\dot{X}$. The topological magnetic charge, $S\mathcal{Q}
$, merely enters the equilibrium background momentum $2\pi S\mathcal{Q}$,
while the collective translation of the kinks with the velocity $\dot{X}$
gives the mass $M$.

From a mathematical viewpoint, however, the starting Lagrangian with the
Berry phase and the axial coupling [see Eq. (\ref{Lagran}) below] is not
manifestly Galilean invariant. Then, the following problems remain highly
non-trivial: (1) how to describe the stability of the ground state under the
Galilean boost, and (2) how to justify the existence of the linear momentum
as a conserved Noether current. To guarantee the stability, Lie point group
symmetries of nonlinear differential equations (DE) describing dynamics of
the system should embrace the whole transformation. To justify the existence
of the conserved current, we need to establish appropriate conservation laws
from the analyses of variational symmetries. The latter plays an important
role in the study of magnon transport. In this paper, to clarify the above
mentioned problems (1) and (2), we start with the continuum version of the
chiral XY model under the magnetic field that describes the chiral
helimagnet with soliton lattice. Then, we apply the Lie group methods and
symmetry analysis to the model. Although the model and the method of the
present analysis have been well established, we are definitely motivated by
how to justify the emergence of the spin current in real chiral helimagnets
that serve as potentially promising devices in spintronics.

The paper is organized as follows. We discuss the model of chiral helimagnet
in Sec. I. In Sec. II we consider the problem of finding the Lie point
symmetries of the DE system of the continuum theory. The main question
studied here is how to find group-invariant solutions of the differential
equations. The variational symmetries are deduced in Sec. II. and we
establish here explicit formulae for the conserved densities and density
currents involved in the conservation laws. Conclusions are given in Sec. IV.

\section{Model}

To describe chiral spin ordering that occurs along a crystallographic axis,
say $z$-axis, the relevant t model Hamiltonian is given by%
\begin{equation}
H=-\mathcal{J}\sum\limits_{j}\vec{S}_{j}\vec{S}_{j+1}-\mathcal{\vec{D}\cdot }%
\sum\limits_{j}\left[ \vec{S}_{j}\times \vec{S}_{j+1}\right] -2\mu
_{0}h\sum\limits_{j}S_{j}^{x}.  \label{Hchir}
\end{equation}%
The ferromagnetic interaction with the exchange coupling $\mathcal{J}$
(first term), and the parity-violating DM interaction, described by the
uniform axial vector $\mathcal{\vec{D}}$, between the nearest-neighbor spins
lead to the long-period incommensurate helimagnetic structure with either
left-handed or right-handed chirality. The direction of the mono-axial $%
\mathcal{\vec{D}}$ vector is fixed by the crystallographic symmetry that we
do not get into the detail\cite{BKO1}. The third term is the Zeeman term
with a transverse magnetic field with its strength being $h$.

In a continuum approximation, we obtain the stationary conditions for the $%
\theta $ and $\varphi $ variables that lead to the coupled nonlinear partial
differential equations given by 
\[
S\,\theta _{t}+\mathcal{J}S^{2}\left( \sin \theta \,\varphi _{zz}+2\cos
\theta \,\theta _{z}\varphi _{z}\right) -2\mathcal{D}S^{2}\cos \theta
\,\theta _{z}-2\mu _{0}hS\sin \varphi =0, 
\]%
\begin{equation}
S\sin \theta \,\varphi _{t}+\mathcal{J}S^{2}\left( \sin \theta \cos \theta
\varphi _{z}^{2}-\theta _{zz}\right) -2\mathcal{D}S^{2}\sin \theta \cos
\theta \varphi _{z}-2\mu _{0}hS\cos \theta \cos \varphi =0.  \label{BEQ0}
\end{equation}%
The energy functional is then given by 
\[
\tilde{H}=\int dz\left\{ \frac{\mathcal{J}S^{2}}{2}\left[ \theta
_{z}^{2}+\sin ^{2}\theta \varphi _{z}^{2}\right] -\mathcal{D}S^{2}\sin
^{2}\theta \varphi _{z}-2\mu _{0}hS\sin \theta \cos \varphi \right\} . 
\]%
To describe the dynamics in an appropriate way, we need the Lagrangian
density,%
\begin{eqnarray}
L &=&S\left( \cos \theta -1\right) \varphi _{t}-\frac{\mathcal{J}S^{2}}{2}%
\left[ \theta _{z}^{2}+\sin ^{2}\theta \varphi _{z}^{2}\right]  \nonumber \\
&&+\mathcal{D}S^{2}\sin ^{2}\theta \varphi _{z}+2\mu _{0}hS\sin \theta \cos
\varphi  \label{Lagran}
\end{eqnarray}%
where the first term is the Berry phase term. The ground state configuration
is then the kink crystal (chiral soliton lattice) state specified by the
stationary solution, 
\begin{equation}
\varphi (z)=2\cos ^{-1}\left[ \mathrm{sn}\left( {\dfrac{m}{\kappa }}z,\kappa
\right) \right] ,\,\,\,\,\,\theta =0,  \label{stationary}
\end{equation}%
where the helical pitch in the zero field is given by $q_{0}=\mathcal{D}/%
\mathcal{J}$, $m=\sqrt{2\mu _{0}h/\mathcal{J}{S}}$, and $\mathrm{sn}$ is the
Jacobi elliptic function with the elliptic modulus $\kappa $ ($0<\kappa <1$)
determined through minimization of energy per unit length, $\kappa /m={%
4E(\kappa )/\pi q_{0}}$, with $E(\kappa )$ being the elliptic integrals of
the second kind.\cite{Dzyaloshinskii} In. Ref. \cite{BKO1}, we constructed
the sliding solution of the kink crystal by introducing the position of the
kink center $Z$ as a dynamical variable. Based on the collective coordinate
method and the Dirac's canonical formulation for the singular Lagrangian
system, we derived the closed formulae for the mass, spin current and
induced magnetic dipole moment accompanied with the kink crystal motion.

\section{Classical Lie symmetries and Galilean boosts}

Now, we perform the classical Lie symmetry analysis to manifest the
existence of the Galilei-boosted solution of the stationary solution (\ref%
{stationary}). The system of the second order DEs, (\ref{BEQ0}), are
eppressed as $F_{1}=0$, $F_{2}=0$, where 
\[
F_{1}=S\sin \theta \,\theta _{t}+\mathcal{J}S^{2}\sin \theta \left( \sin
\theta \,\varphi _{zz}+2\cos \theta \,\theta _{z}\varphi _{z}\right) -%
\mathcal{D}S^{2}\sin 2\theta \,\theta _{z}-2\mu _{0}hS\sin \theta \sin
\varphi , 
\]%
\begin{equation}
F_{2}=S\sin \theta \,\varphi _{t}+\mathcal{J}S^{2}\left( \sin \theta \cos
\theta \varphi _{z}^{2}-\theta _{zz}\right) -2\mathcal{D}S^{2}\sin \theta
\cos \theta \varphi _{z}-2\mu _{0}hS\cos \theta \cos \varphi ,  \label{BEQ}
\end{equation}%
with two dependent variables $u=(\theta ,\varphi )$, and two independent
variables $w=(z,t)$. The first equation in (\ref{BEQ}) is multiplied by the
factor $\sin \theta $ that is necessary for the further symmetry analysis.
The symmetry condition requires that the system must hold in the transformed
variables whenever it holds in the original variables.

In order to take into account the derivative terms involved in Eq.(\ref{BEQ}%
), the infinitesimal generator is the prolongation introduced as\cite{Olver}
($i=1$ means $"z"$, $i=2$ does $"t"$) 
\[
\underset{2}{\hat{X}}=\xi \frac{\partial }{\partial z}+\eta \frac{\partial }{%
\partial t}+\zeta \frac{\partial }{\partial \theta }+\chi \frac{\partial }{%
\partial \varphi }+\sum\limits_{i=1}^{2}\zeta _{i}\frac{\partial }{\partial
\theta _{i}}+\sum\limits_{i=1}^{2}\chi _{i}\frac{\partial }{\partial \varphi
_{i}}+\sum_{\underset{{\large (i\leq j)}}{i,j=1}\,}^{2}\;\zeta _{ij}\frac{%
\partial }{\partial \theta _{ij}}+\sum_{\underset{{\large (i\leq j)}}{i,j=1}%
\,}^{2}\;\chi _{ij}\frac{\partial }{\partial \varphi _{ij}},
\]%
where the first prolongation is given by 
\[
\zeta _{i}=D_{i}\left( \zeta \right) -\theta _{z}D_{i}\left( \xi \right)
-\theta _{t}D_{i}\left( \eta \right) ,\;\chi _{i}=D_{i}\left( \chi \right)
-\varphi _{z}D_{i}\left( \xi \right) -\varphi _{t}D_{i}\left( \eta \right)
,\;(i=z,t)
\]%
and the second one is determined from the first ones 
\[
\zeta _{1j}=D_{j}\left( \zeta _{1}\right) -\theta _{zz}D_{j}\left( \xi
\right) -\theta _{zt}D_{j}\left( \eta \right) ,\;\chi _{1j}=D_{j}\left( \chi
_{1}\right) -\varphi _{zz}D_{j}\left( \xi \right) -\varphi _{zt}D_{j}\left(
\eta \right) ,
\]%
\[
\zeta _{2j}=D_{j}\left( \zeta _{2}\right) -\theta _{zt}D_{j}\left( \xi
\right) -\theta _{tt}D_{j}\left( \eta \right) ,\;\chi _{2j}=D_{j}\left( \chi
_{2}\right) -\varphi _{zt}D_{j}\left( \xi \right) -\varphi _{tt}D_{j}\left(
\eta \right) .
\]%
Here, the $D_{i}$ are the total derivatives with respect to $z$ and $t$
relevant for the system (\ref{BEQ}).

The system of DEs is invariant under the group $G$ if 
\begin{equation}
\underset{2}{\hat{X}}F_{i}|_{F}=0,\;i=1,2.  \label{invar}
\end{equation}%
The invariance condition provides a partial differential equations involving
the unknown infinitesimal functions with their derivatives and products of
the partial derivatives of the dependent variables. After replacing $\theta
_{zz}$ and $\varphi _{zz}$ from Eq.(\ref{BEQ}) and splitting the resulting
equation into powers of derivatives of the dependent variables (shown below
at left), handled as independent ones, the overdetermined system $\underset{2%
}{\hat{X}}F_{1}|_{F}=0$ is obtained 
\begin{equation}
\varphi _{zt}:\sin ^{2}\theta \,\eta _{z}=0  \label{de1}
\end{equation}%
\begin{equation}
\varphi _{zt}\theta _{z}:\sin ^{2}\theta \,\eta _{\theta }=0  \label{de2}
\end{equation}%
\begin{equation}
\varphi _{zt}\varphi _{z}:\sin ^{2}\theta \,\eta _{\varphi }=0  \label{de3}
\end{equation}%
\begin{equation}
\theta _{z}\theta _{t}:\xi _{\theta }\sin \theta =0  \label{de4}
\end{equation}%
\begin{equation}
\theta _{z}\varphi _{t}:\xi _{\varphi }\sin \theta =0  \label{de5}
\end{equation}%
\begin{equation}
\theta _{t}:-\zeta \cos \theta -\eta _{t}\sin \theta +2\xi _{z}\sin \theta
-\chi _{\varphi }\sin \theta +\zeta _{\theta }\sin \theta =0  \label{de6}
\end{equation}%
\begin{equation}
\varphi _{t}:\zeta _{\varphi }\sin \theta +\chi _{\theta }\sin ^{3}\theta =0
\label{de7}
\end{equation}%
\begin{equation}
\theta _{z}\varphi _{z}:-\chi _{\theta \varphi }+2\zeta +\chi _{\theta
\varphi }\cos 2\theta -\zeta _{\theta }\sin 2\theta =0  \label{de8}
\end{equation}%
\begin{equation}
\theta _{z}^{2}:2\chi _{\theta }\cos \theta +\chi _{\theta \theta }\sin
\theta =0  \label{de9}
\end{equation}%
\begin{equation}
\varphi _{z}^{2}:\chi _{\varphi \varphi }\sin ^{2}\theta +\chi _{\theta
}\cos \theta \sin ^{3}\theta +\zeta _{\varphi }\sin 2\theta =0  \label{de10}
\end{equation}%
\[
\theta _{z}:-\mathcal{J}S\chi _{\theta z}\cos 2\theta -\xi _{t}\sin \theta +%
\mathcal{J}S\chi _{\theta z}+2\mathcal{D}S\zeta +\mathcal{J}S\chi _{z}\sin
2\theta 
\]%
\begin{equation}
-\mathcal{D}S\sin 2\theta \left( \xi _{z}-\chi _{\varphi }+\zeta _{\theta
}\right) =0  \label{de11}
\end{equation}%
\begin{equation}
\varphi _{z}:-\mathcal{J}S\sin ^{2}\theta \xi _{zz}+2\mathcal{J}S\sin
^{2}\theta \chi _{\varphi z}-2\mathcal{D}S\cos \theta \sin ^{3}\theta \chi
_{\theta }+\mathcal{J}S\sin 2\theta \zeta _{z}-\mathcal{D}S\sin 2\theta
\zeta _{\varphi }=0  \label{de12}
\end{equation}%
\[
1:\zeta _{t}\sin \theta -2\mu _{0}h\cos \varphi \sin \theta \chi +\mathcal{J}%
S\sin ^{2}\theta \chi _{zz}-2\mu _{0}h\sin ^{2}\theta \cos \theta \cos
\varphi \chi _{\theta } 
\]%
\begin{equation}
-\mathcal{D}S\sin 2\theta \zeta _{z}+2\mu _{0}h\cos \theta \sin \varphi
\zeta -4\mu _{0}h\sin \theta \sin \varphi \xi _{z}+2\mu _{0}h\sin \theta
\sin \varphi \chi _{\varphi }=0.  \label{de13}
\end{equation}%
The overdetermined system $\underset{2}{\hat{X}}F_{2}|_{F}=0$ is written as 
\begin{equation}
\theta _{zt}:\eta _{z}=0  \label{ed1}
\end{equation}%
\begin{equation}
\theta _{zt}\theta _{z}:\eta _{\theta }=0  \label{ed2}
\end{equation}%
\begin{equation}
\theta _{zt}\varphi _{z}:\eta _{\varphi }=0  \label{ed3}
\end{equation}%
\begin{equation}
\varphi _{z}\varphi _{t}:\xi _{\varphi }\sin \theta =0  \label{ed4}
\end{equation}%
\begin{equation}
\varphi _{z}\theta _{t}:\xi _{\theta }\sin \theta =0  \label{ed5}
\end{equation}%
\begin{equation}
\varphi _{t}:\zeta \cos \theta -\eta _{t}\sin \theta +2\xi _{z}\sin \theta
+\chi _{\varphi }\sin \theta -\zeta _{\theta }\sin \theta =0  \label{ed6}
\end{equation}%
\begin{equation}
\theta _{t}:\zeta _{\varphi }+\chi _{\theta }\sin ^{2}\theta =0  \label{ed7}
\end{equation}%
\begin{equation}
\theta _{z}\varphi _{z}:\zeta _{\theta \varphi }-\zeta _{\varphi }\cot
\theta -\chi _{\theta }\cos \theta \sin \theta =0  \label{ed8}
\end{equation}%
\begin{equation}
\theta _{z}^{2}:\zeta _{\theta \theta }=0  \label{ed9}
\end{equation}%
\begin{equation}
\varphi _{z}^{2}:2\zeta _{\varphi \varphi }-2\zeta \cos 2\theta +(\zeta
_{\theta }-2\chi _{\varphi })\sin 2\theta =0  \label{ed10}
\end{equation}%
\[
\varphi _{z}:-2\mathcal{J}S\zeta _{\varphi z}-2\mathcal{D}S\cos 2\theta
\zeta -\xi _{t}\sin \theta -\mathcal{D}S\sin 2\theta \xi _{z}+\mathcal{J}%
S\sin 2\theta \chi _{z} 
\]%
\begin{equation}
-\mathcal{D}S\sin 2\theta \chi _{\varphi }+\mathcal{D}S\sin 2\theta \zeta
_{\theta }=0  \label{ed11}
\end{equation}%
\begin{equation}
\theta _{z}:\mathcal{J}\left( \xi _{zz}-2\zeta _{\theta z}\right) -2\mathcal{%
D}\cot \theta \zeta _{\varphi }-\mathcal{D}\sin 2\theta \chi _{\theta }=0
\label{ed12}
\end{equation}%
\[
1:-\mathcal{J}S\zeta _{zz}-4\mu _{0}h\cos \theta \cos \varphi \xi _{z}+2\mu
_{0}h\cos \theta \cos \varphi \zeta _{\theta }+\chi _{t}\sin \theta +2\mu
_{0}h\sin \theta \cos \varphi \zeta 
\]%
\begin{equation}
-\mathcal{D}S\sin 2\theta \chi _{z}+2\mu _{0}h\cos \theta \sin \varphi \chi
-2\mu _{0}h\frac{\sin \varphi }{\sin \theta }\zeta _{\varphi }=0
\label{ed13}
\end{equation}

The general solution of the determining equations is presented in Appendix
A. The final result is 
\begin{equation}
\xi =C_{1},\;\eta =C_{2},\;\chi =0,\;\zeta =0,  \label{gensol}
\end{equation}%
where $C_{1,2}$ are the constants. Therefore, every infinitesimal generator
of Lie point symmetries is a linear combination of $\hat{X}_{1}=\partial
_{z} $\ and $\hat{X}_{2}=\partial _{t}$%
\begin{equation}
\hat{X}=C_{1}\partial _{z}+C_{2}\partial _{t}.  \label{LieAlgeb}
\end{equation}%
We seek invariant solutions of the Eqs.(\ref{BEQ}) under symmetries
generated by (\ref{LieAlgeb}). The characteristic equations are

\[
\frac{dz}{C_1}=\frac{dt}{C_2}=\frac{d\theta }0=\frac{d\varphi }0, 
\]
which have three functionally independent first integrals 
\[
\mathcal{I}_1=z-vt,\;\mathcal{I}_2=\theta ,\;\mathcal{I}_3=\varphi , 
\]
where $v=C_1/C_2$ is a velocity, and $\hat{X}\mathcal{I}_i=0$ ($i=1,2,3$).

We can insert first integrals having the form 
\begin{equation}
\theta =\theta \left[ z-vt\right] ,\;\varphi =\varphi \left[ z-vt\right]
\label{travel}
\end{equation}%
into the original system (\ref{BEQ}). This yields the simplified partial DE
system after introducing the new coordinate $\tilde{z}=z-vt$. Thus, the
invariant solution includes the sliding mode (\ref{travel}) which can be
found by inspection, but we classify \textit{all} invariant solutions, using
the structure of the Lie algebra. Therefore, we proved that the
Galilei-boosted solution certainly exists.

\section{Variational symmetries and conservation laws}

Next, we construct the conserved Noether current to justify the existence of
the linear momentum that eventually carries the spin current in chiral
helimagnets. A derivation of conservation laws is based on so-called Noether
identity\cite{Ibragimov} 
\begin{equation}
\hat{Y}\left( L\right) +\hat{D}_{i}\cdot \left( \xi ^{i}L\right) =W^{\alpha
}E^{\alpha }(L)+D_{i}\hat{N}^{i}(L),\;\left( \alpha =1,2;\,i=z,t\right)
\label{Noether}
\end{equation}%
where the canonical vector field eligible for the Lagrangian (\ref{Lagran}) 
\[
\hat{Y}=W^{1}\partial _{\theta }+W^{2}\partial _{\varphi }+D_{z}\left(
W^{1}\right) \partial _{\theta _{z}}+D_{z}\left( W^{2}\right) \partial
_{\varphi _{z}}+D_{t}\left( W^{2}\right) \partial _{\varphi _{t}}, 
\]%
with the characteristics 
\[
W^{1}=\eta ^{1}-\xi ^{t}\theta _{t}-\xi ^{z}\theta _{z},\;W^{2}=\eta
^{2}-\xi ^{t}\varphi _{t}-\xi ^{z}\varphi _{z}. 
\]%
The dot in Eq. (\ref{Noether}) means the differentiation rule $\hat{D}%
_{i}\cdot \left( \xi ^{i}L\right) \equiv \hat{D}_{i}\left( \xi ^{i}\right)
L+\xi ^{i}\hat{D}_{i}\left( L\right) $.

In Eq. (\ref{Noether}) $N^{i}$ are the Noether operators given by the
expressions 
\[
\hat{N}^{i}=\xi ^{i}+\left( \eta ^{\alpha }-\xi ^{l}u_{l}^{\alpha }\right)
\left( \frac{\partial }{\partial u_{i}^{\alpha }}+\sum\limits_{s\geq
1}\left( -1\right) ^{s}D_{j_{1}}\ldots D_{j_{s}}\frac{\partial }{\partial
u_{ij_{1}\ldots j_{s}}^{\alpha }}\right) 
\]%
\begin{equation}
+\sum\limits_{r\geq 1}\left( -1\right) ^{r}D_{k_{1}}\ldots D_{k_{r}}\left(
\eta ^{\alpha }-\xi ^{l}u_{l}^{\alpha }\right) \left( \frac{\partial }{%
\partial u_{ik_{1}\ldots k_{r}}^{\alpha }}+\sum\limits_{s\geq 1}\left(
-1\right) ^{s}D_{j_{1}}\ldots D_{j_{s}}\frac{\partial }{\partial
u_{ik_{1}\ldots k_{r}j_{1}\ldots j_{s}}^{\alpha }}\right) ,  \label{Noper}
\end{equation}%
and Euler-Lagrange operators are 
\[
E^{\alpha }=\frac{\partial }{\partial u^{\alpha }}+\sum\limits_{s\geq
1}\left( -1\right) ^{s}D_{i_{1}}\ldots D_{i_{s}}\frac{\partial }{\partial
u_{i_{1}\ldots i_{s}}^{\alpha }}. 
\]%
The variational symmetries on the Euler-Lagrange equations may be found via
the Noether identity 
\[
\hat{Y}\left( L\right) +D_{z}\cdot \left( \xi ^{z}L\right) +D_{t}\cdot
\left( \xi ^{t}L\right) =0. 
\]%
In explicit form it is written as 
\[
0=\eta ^{1}L_{\theta }+\eta ^{2}L_{\varphi }+\left( D_{z}\left( W^{1}\right)
+\xi ^{t}\theta _{zt}+\xi ^{z}\theta _{zz}\right) L_{\theta _{z}}+\left(
D_{z}\left( W^{2}\right) +\xi ^{t}\varphi _{zt}+\xi ^{z}\varphi _{zz}\right)
L_{\varphi _{z}} 
\]%
\[
+\left( D_{t}\left( W^{2}\right) +\xi ^{t}\varphi _{tt}+\xi ^{z}\varphi
_{zt}\right) L_{\varphi _{t}}+\left( D_{z}\left( \xi ^{z}\right)
+D_{t}\left( \xi ^{t}\right) \right) L 
\]%
with the derivatives 
\[
L_{\theta }=-S\sin \theta \varphi _{t}-\frac{\mathcal{J}S^{2}}{2}\sin
2\theta \varphi _{z}^{2}+\mathcal{D}S^{2}\sin 2\theta \,\varphi _{z}+2\mu
_{0}hS\cos \theta \cos \varphi , 
\]%
\[
L_{\varphi }=-2\mu _{0}hS\sin \theta \sin \varphi ,\;L_{\theta _{z}}=-%
\mathcal{J}S^{2}\theta _{z}, 
\]%
\[
L_{\varphi _{z}}=-\mathcal{J}S^{2}\sin ^{2}\theta \varphi _{z}^{2}+\mathcal{D%
}S^{2}\sin ^{2}\theta ,\;L_{\varphi _{t}}=S\left( \cos \theta -1\right) . 
\]%
After a little manipulation one gets the following equation for the unknowns 
\[
0=\eta ^{1}\left( -S\sin \theta \,\varphi _{t}-\frac{\mathcal{J}S^{2}}{2}%
\sin 2\theta \varphi _{z}^{2}+\mathcal{D}S^{2}\sin 2\theta \,\varphi
_{z}+2\mu _{0}hS\cos \theta \cos \varphi \right) 
\]%
\[
-\eta ^{2}\,2\mu _{0}hS\sin \theta \sin \varphi 
\]%
\[
-\mathcal{J}S^{2}\theta _{z}\left( \eta _{z}^{1}+\theta _{z}\eta _{\theta
}^{1}+\varphi _{z}\eta _{\varphi }^{1}-\theta _{t}\left[ \xi _{z}^{t}+\theta
_{z}\xi _{\theta }^{t}+\varphi _{z}\xi _{\varphi }^{t}\right] -\theta _{z}%
\left[ \xi _{z}^{z}+\theta _{z}\xi _{\theta }^{z}+\varphi _{z}\xi _{\varphi
}^{z}\right] \right) 
\]%
\[
-\left( \mathcal{J}S^{2}\sin ^{2}\theta \varphi _{z}-\mathcal{D}S^{2}\sin
^{2}\theta \right) 
\]%
\[
\times \left( \eta _{z}^{2}+\theta _{z}\eta _{\theta }^{2}+\varphi _{z}\eta
_{\varphi }^{2}-\varphi _{t}\left[ \xi _{z}^{t}+\theta _{z}\xi _{\theta
}^{t}+\varphi _{z}\xi _{\varphi }^{t}\right] -\varphi _{z}\left[ \xi
_{z}^{z}+\theta _{z}\xi _{\theta }^{z}+\varphi _{z}\xi _{\varphi }^{z}\right]
\right) 
\]%
\[
+S\left( \cos \theta -1\right) \left( \eta _{t}^{2}+\theta _{t}\eta _{\theta
}^{2}+\varphi _{t}\eta _{\varphi }^{2}-\varphi _{t}\left[ \xi
_{t}^{t}+\theta _{t}\xi _{\theta }^{t}+\varphi _{t}\xi _{\varphi }^{t}\right]
-\varphi _{z}\left[ \xi _{t}^{z}+\theta _{t}\xi _{\theta }^{t}+\varphi
_{t}\xi _{\varphi }^{t}\right] \right) 
\]%
\[
+\left( S\left( \cos \theta -1\right) \varphi _{t}-\frac{\mathcal{J}S^{2}}{2}%
\left( \theta _{z}^{2}+\sin ^{2}\theta \varphi _{z}^{2}\right) +\mathcal{D}%
S^{2}\sin ^{2}\theta \,\varphi _{z}+2\mu _{0}hS\sin \theta \cos \varphi
\right) 
\]%
\[
\times \left( \xi _{t}^{t}+\theta _{t}\xi _{\theta }^{t}+\varphi _{t}\xi
_{\varphi }^{t}+\xi _{z}^{z}+\theta _{z}\xi _{\theta }^{z}+\varphi _{z}\xi
_{\varphi }^{z}\right) . 
\]%
The equation is solved as an algebraic equation with respect to the partial
derivatives of the dependent variables, handled as independent variables.
The equation leads to a determining equations with respect to the unknown
infinitesimal functions $\xi ^{z}$, $\xi ^{t}$, $\eta ^{1}$ and $\eta ^{2}$.
The details of the calculations are relegated into Appendix B. As a result,
we obtain, $C_{1,2}$ are constants, 
\begin{equation}
\xi ^{t}=C_{1},\;\xi ^{z}=C_{2},\;\eta ^{1}=\eta ^{2}=0.  \label{varsol}
\end{equation}%
Using the found variational symmetries the operators $\hat{N}^{i}$ are
firstly constructed from formulae (\ref{Noper}) and the corresponding
conservation laws are then calculated from 
\[
D_{i}\left( \hat{N}^{i}L\right) =0. 
\]%
The Noether operators modify the Lagrangian density into a conserved
quantity $\mathcal{C}^{i}=N^{i}(L)$ with a zero total divergence.

(i) The choice $C_1=1$ and $C_2=0$ corresponds to the symmetry group under
translations in time $t\rightarrow t+a$. Then 
\[
\hat{N}^t=1-\theta _t\frac \partial {\partial \theta _t}-\varphi _t\frac
\partial {\partial \varphi _t},\;\hat{N}^z=-\theta _t\frac \partial
{\partial \theta _z}-\varphi _t\frac \partial {\partial \varphi _z}. 
\]
By introducing the functions 
\[
\mathcal{C}^t=N^t(L)=-\frac{\mathcal{J}S^2}2\sin ^2\theta \,\varphi _z^2-%
\frac{\mathcal{J}S^2}2\theta _z^2+\mathcal{D}S^2\sin ^2\theta \,\varphi
_z+2\mu _0hS\sin \theta \cos \varphi , 
\]
and 
\[
\mathcal{C}^z=N^z(L)=\mathcal{J}S^2\,\theta _z\theta _t+\mathcal{J}S^2\sin
^2\theta \,\varphi _z\varphi _t-\mathcal{D}S^2\sin ^2\theta \,\,\varphi _t. 
\]
one gets the energy conservation law $D_t\left( \mathcal{C}^t\right)
+D_z\left( \mathcal{C}^z\right) =0$, or 
\[
D_t\left( \mathcal{E}\right) =D_z\left( J_z^{(\mathcal{E})}\right) , 
\]
where $\mathcal{E}=-\mathcal{C}^t$ and $J_z^{(\mathcal{E})}=\mathcal{C}^z$
are the energy density and the density of energy current, respectively.

(ii) The choice $C_{1}=0$ and $C_{2}=1$ yields the variational symmetry
under translations in space $z\rightarrow z+a$. Then 
\[
\hat{N}^{t}=-\theta _{z}\frac{\partial }{\partial \theta _{t}}-\varphi _{z}%
\frac{\partial }{\partial \varphi _{t}},\;\hat{N}^{z}=1-\theta _{z}\frac{%
\partial }{\partial \theta _{z}}-\varphi _{z}\frac{\partial }{\partial
\varphi _{z}}. 
\]%
By the same manner one can obtain the momentum conservation law 
\[
D_{t}\left( P_{z}\right) =D_{z}\left( T_{zz}\right) , 
\]%
with the momentum $P_{z}=-\mathcal{C}^{t}=-N^{t}(L)$%
\[
P_{z}=S\left( 1-\cos \theta \right) \varphi _{z}, 
\]%
and the canonical energy-momentum tensor $T_{zz}=\mathcal{C}^{z}=N^{z}(L)$%
\[
T_{zz}=S\left( \cos \theta -1\right) \varphi _{t}+\frac{\mathcal{J}S^{2}}{2}%
\left( \theta _{z}^{2}+\sin ^{2}\theta \,\varphi _{z}^{2}\right) +2\mu
_{0}hS\sin \theta \cos \varphi . 
\]%
Therefore we proved the existence of the linear momentum as a conserved
Noether current. Then, we are ready to define the longitudinal spin current
carried by the kink crystal by writing the linear momentum per unit area as%
\begin{equation}
\mathcal{P}_{z}=S\int_{0}^{L_{0}}\left( 1-\cos \theta \right) \varphi
_{z}\,dz=2\pi S\mathcal{Q}+M\dot{z},  \label{momentum_per_area}
\end{equation}%
where $L_{0}$ is the system size. The topological magnetic charge, $S%
\mathcal{Q} $, merely produces the equilibrium background momentum $2\pi S%
\mathcal{Q}$, while the collective translation of the kinks with the
velocity $\dot{z}$ gives the inertial mass $M$ of the kink crystal\cite{BKO1}%
.

\section{Conclusions}

In this paper, motivated by the spin current problem in chiral helimagnet,
we rigorously proved the hidden Galilean invariance embedded in the chiral
XY model under the magnetic field. The Lie group analysis is applied to the
differential equations of the continuum theory of the chiral helimagnet with
the parity-violating Dzyaloshinskii-Morya coupling under a transversal
magnetic field. Lie point symmetries and the invariant solutions under these
symmetries are found. They present sliding solutions that come up as a
consequence of both a breaking of spin rotational symmetry by the external
magnetic field and a parity violation due to DM interaction. We found that
variational symmetries are related with translations in space and time, the
corresponding energy and momentum conservation laws are derived. We
therefore succeeded in justifying the existence of the transport spin
current in chiral helimagnet.\bigskip 

\appendix\textbf{Appendix A}

From Eqs.(\ref{de1}-\ref{de5}) and (\ref{ed1}-\ref{ed5}) one obtain $\eta
=\eta (t)$ and $\xi =\xi (z,t)$. From Eq.(\ref{de9}) we get $\chi =\chi
_0(z,t,\varphi )-\chi _1(z,t,\varphi )\cot \theta $ and $\zeta =\zeta
_1(z,t,\varphi )\theta +\zeta _0(z,t,\varphi )$ from Eq.(\ref{ed9}).

(i) Let us consider $\chi =\chi _0(z,t,\varphi )$. A substitution of the
results for $\chi $ and $\zeta $ into Eq.(\ref{de6}) yields 
\[
-\left( \zeta _1\theta +\zeta _0\right) \cos \theta -\eta _t\sin \theta
+2\xi _z\sin \theta -\chi _{0\varphi }\sin \theta +\zeta _1\sin \theta =0 
\]
and after splitting over the $\theta $ variable we obtain $\zeta =0$ ($\zeta
_0=\zeta _1=0$) and 
\begin{equation}
-\eta _t+2\xi _z-\chi _{0\varphi }=0.  \label{newEq}
\end{equation}
From Eq.(\ref{de7}) we get $\chi _\theta =0$ that agrees with the choice $%
\chi =\chi _0(z,t,\varphi )$ and transforms (\ref{ed8}) into identity. A sum
of Eqs.(\ref{de6}) and (\ref{ed6}) gives $\eta _t=2\xi _z$, and, therefore, $%
\xi _{zz}=0$ and $\chi _{0\varphi }=0$ (from Eq.[\ref{newEq}]), i.e. $\chi
=\chi _0(z,t)$. Then Eqs.(\ref{de10},\ref{ed10}) become identities. Eqs.(\ref%
{de11},\ref{ed11}) coincide with each other and may be written as 
\[
-\xi _t\sin \theta -\mathcal{D}S\sin 2\theta \xi _z+\mathcal{J}S\sin 2\theta
\chi _z=0. 
\]
The relation splits into 
\[
\xi _t=0,\;-\mathcal{D}\xi _z+\mathcal{J}\chi _z=0, 
\]
that gives $\chi _{zz}=0$. Eqs.(\ref{de12},\ref{ed12}) are identically
fulfilled. The last Eqs.(\ref{de13},\ref{ed13}) take the form 
\[
\cos \varphi \sin \theta \chi +2\sin \theta \sin \varphi \xi _z=0, 
\]
\[
-4\mu _0h\cos \theta \cos \varphi \xi _z+\chi _t\sin \theta -\mathcal{D}%
S\sin 2\theta \chi _z+2\mu _0h\cos \theta \sin \varphi \chi =0, 
\]
respectively. Since $\chi $ does not depend on $\theta $ and $\varphi $,
after splitting over these variables we get 
\[
\chi _t=0,\;\chi _z=0,\;\chi =0,\;\xi _z=0. 
\]
As a result we have eventually $\xi =$const$,\;\eta =$const$,\;\chi
=0,\;\zeta =0.$

(ii) Now we take $\chi =-\chi _{1}(z,t,\varphi )\cot \theta $. We prove that
only $\chi _{1}=0$ satisfies the determining equations. Indeed, a
substitution of $\chi $ into Eq.(\ref{ed6}) and a splitting over $\theta $
variable gives 
\[
\zeta _{1}=0,\;\zeta _{0}=\chi _{1\varphi },\;\eta _{t}=2\xi _{z}.
\]%
This means that $\zeta =\zeta _{0}(z,t,\varphi )$ and $\xi _{zz}=0$, and Eq.(%
\ref{de6}) becomes identity. From Eq.(\ref{ed7}) we obtain $\zeta _{0\varphi
}=-\chi _{1}$ that results together with $\zeta _{0}=\chi _{1\varphi }$ in 
\[
\chi =-\left[ a(z,t)\cos \varphi +b(z,t)\sin \varphi \right] \cot \theta ,
\]%
\[
\zeta =-a(z,t)\sin \varphi +b(z,t)\cos \varphi .
\]%
Eqs.(\ref{de7},\ref{de8},\ref{ed8},\ref{de10},\ref{ed10},\ref{de12},\ref%
{ed12}) are identically fulfilled. After splitting over $\theta $ and $%
\varphi $ variables Eq.(\ref{ed11}) turns into 
\[
\sin \theta :\;\xi _{t}=0,\;\;\;\;\sin 2\theta :\;\xi _{z}=0,
\]%
\[
\cos \varphi :\;\mathcal{J}a_{z}+\mathcal{D}b=0,\;\;\;\;\sin \varphi :\;%
\mathcal{J}b_{z}-\mathcal{D}a=0,
\]%
\[
\cos ^{2}\theta \cos \varphi :\;\mathcal{D}b+\mathcal{J}a_{z}=0,\;\;\;\;\cos
^{2}\theta \sin \varphi :\;-\mathcal{D}a+\mathcal{J}b_{z}=0,
\]%
that yields $\xi =$const, $\eta =$const, and $a_{z}=-\left( \mathcal{D}/%
\mathcal{J}\right) b$, $b_{z}=\left( \mathcal{D}/\mathcal{J}\right) a$. Eq.(%
\ref{de11}) gives the same result. From the last Eq.(\ref{de13}) we obtain 
\[
\sin \theta \sin \varphi :\;a_{t}=0,\;\;\;\;\sin \theta \cos \varphi
:\;b_{t}=0,
\]%
i.e. $a=a(z)$ and $b=b(z)$ and 
\[
\sin \theta \cos \theta \cos \varphi :\;\mathcal{J}a_{zz}+2\mathcal{D}%
b_{z}=0,
\]%
\[
\sin \theta \cos \theta \sin \varphi :\;\mathcal{J}b_{zz}-2\mathcal{D}%
a_{z}=0.
\]%
Together with the previous results these equations give $a_{z}=0$ and $%
b_{z}=0$, therefore $a=b=0$, and we get eventually $\chi =\zeta =0$. Then
Eq.(\ref{ed13}) will be an identity. Eventually, the result (\ref{gensol})
is again obtained.

\newpage 

\appendix\textbf{Appendix B}

A set of equations 
\[
\begin{array}{cc}
\theta _z^3: & \xi _\theta ^z=0 \\ 
\theta _z^2\varphi _z: & \xi _\varphi ^z=0 \\ 
\theta _z^2\theta _t: & \xi _\theta ^t=0 \\ 
\theta _z\theta _t\varphi _z: & \xi _\varphi ^t=0%
\end{array}
\]
yields immediately $\xi ^z=\xi ^z(z,t)$ and $\xi ^t=\xi ^t(z,t)$. A
remaining part acquires the form 
\begin{equation}
\begin{tabular}{ll}
$\theta _z\theta _t:$ & $\xi _z^t=0$%
\end{tabular}
\label{first}
\end{equation}
\begin{equation}
\begin{array}{cc}
\theta _z^2: & \eta _\theta ^1-\frac 12\xi _z^z+\frac 12\xi _t^t=0%
\end{array}
\label{second}
\end{equation}
\begin{equation}
\begin{array}{cc}
\theta _z\varphi _z: & \eta _\varphi ^1+\sin ^2\theta \eta _\theta ^2=0%
\end{array}
\label{third}
\end{equation}
\begin{equation}
\begin{array}{cc}
\varphi _z^2: & -\frac 12\sin 2\theta \,\eta ^1+\sin ^2\theta \,\xi
_z^z-\sin ^2\theta \,\eta _\varphi ^2-\frac 12\sin ^2\theta \,\left( \xi
_t^t+\xi _z^z\right) =0%
\end{array}
\label{fourth}
\end{equation}
\begin{equation}
\begin{array}{cc}
\theta _t: & (\cos \theta -1)\eta _\theta ^2=0\;\Longrightarrow \eta _\theta
^2=0%
\end{array}
\label{fifth}
\end{equation}
\begin{equation}
\begin{array}{cc}
\theta _z: & -\mathcal{J}\eta _z^1+\mathcal{D}\sin ^2\theta \,\eta _\theta
^2=0%
\end{array}
\label{sixth}
\end{equation}
\begin{equation}
\begin{array}{cc}
\varphi _t: & -\sin \theta \,\eta ^1+\left( \cos \theta -1\right) \,\left(
\eta _\varphi ^2+\xi _z^z\right) =0%
\end{array}
\label{seventh}
\end{equation}
\[
\begin{array}{cc}
\varphi _z: & \mathcal{D}S^2\sin 2\theta \,\,\eta ^1-\mathcal{J}S^2\sin
^2\theta \eta _z^2+\mathcal{D}S^2\sin ^2\theta \eta _\varphi ^2%
\end{array}
\]

\begin{equation}
\begin{array}{cc}
& -\mathcal{D}S^2\sin ^2\theta \xi _z^z-S\left( \cos \theta -1\right) \,\xi
_t^z+\mathcal{D}S^2\sin ^2\theta \left( \xi _t^t+\xi _z^z\right) =0%
\end{array}
\label{eighth}
\end{equation}

\[
\begin{array}{cc}
1: & 2\mu _{0}hS\cos \theta \cos \varphi \,\eta ^{1}-2\mu _{0}hS\sin \theta
\sin \varphi \,\eta ^{2}+\mathcal{D}S^{2}\sin ^{2}\theta \,\eta _{z}^{2}%
\end{array}%
\]%
\begin{equation}
\begin{array}{cc}
& +S\left( \cos \theta -1\right) \eta _{t}^{2}+2\mu _{0}hS\sin \theta \cos
\varphi \,\left( \xi _{t}^{t}+\xi _{z}^{z}\right) =0%
\end{array}
\label{nineth}
\end{equation}%
From Eqs.(\ref{first}) and (\ref{fifth}) we obtain $\xi ^{t}=\xi ^{t}(t)$
and $\eta ^{2}=\eta ^{2}\left( z,t,\varphi \right) $. Together with Eqs.(\ref%
{third}) and (\ref{sixth}) we get $\eta ^{1}=\eta ^{1}\left( t,\theta
\right) $. From Eq.(\ref{second}) $\eta _{\theta \theta }^{1}=0$, hence $%
\eta ^{1}=a(t)\theta +b(t)$. A substitution of the result into Eq.(\ref%
{seventh}) yields $a=0$, $b=0$ and $\eta _{\varphi }^{2}+\xi _{z}^{z}=0$.
This means $\eta ^{1}=0$ and $\xi _{t}^{t}=\xi _{z}^{z}$ [Eq.(\ref{second}%
)]. Together with Eq.(\ref{fourth}) this produces $\eta _{\varphi }^{2}=0$
and $\xi _{z}^{z}=\xi _{t}^{t}=0$, i.e. $\xi ^{t}=$const and $\xi ^{z}=\xi
^{z}(t)$. Using Eq.(\ref{eighth}) we get after splitting over $\theta $
variable $\xi ^{z}=$const and $\eta ^{2}=\eta ^{2}\left( t\right) $. It
means that $\eta _{z}^{2}=0$ and we get $\eta ^{2}=0$ from Eq.(\ref{nineth}%
). By collecting all results together we obtain (\ref{varsol}).

\end{document}